\pdfoutput=1
\documentclass[amssymb,amsmath]{article} 
\usepackage[english]{babel}
\usepackage{graphics}
\usepackage{graphicx}
\usepackage{parskip}
\DeclareGraphicsExtensions{.pdf,.png,.jpg}
\usepackage{cite}
\title{A compact theoretical model for opto-electronic devices based on quantum dot arrays}
\author{S. Illera, J. D. Prades, A. Cirera and A. Cornet\\
\small MIND/IN$^2$UB Departament d'Electr\`onica, Universitat de Barcelona,\\\small C/Mart\'i i Franqu\`es 1, E-08028 Barcelona, Spain}
\date{}
\begin{document}
\maketitle

\hrulefill
\begin{abstract}
We present a theoretical model describing the photo-electrical response of a system composed of quantum dots embedded in a dielectric matrix. The model is based on the non-coherent rate equations and the Transfer Hamiltonian formalism. The self charge was included. Within this methodology, the response of the system only depends on fundamental material parameters and its geometry.\\
Transport through several quantum dot configurations was simulated obtaining current-voltage curves in dark and illuminating conditions for three different scenarios: single one quantum dot and two quantum dots in parallel and serial configurations. Despite the simplicity of the model, it has been used to reproduce successfully experimental results.
\end{abstract}
% insert suggested PACS numbers in braces on next line
\small Contact author: sillera@el.ub.edu

\section{Introduction}
Silicon quantum dots (Si Qd) embedded in insulator matrices have opened a new branch of possibilities in the electronics based on quantum confinement and Coulomb blockade effects \cite{propQd, Cblock}. Compared to the standard bulk Si technology, devices based in these structures have increased the fabrication and conceptual complexity. Qd can be used in single-electron transistors \cite{Meir}, new memory concepts \cite{tiwari} and photon or electroluminescent devices \cite{lock}. Concerning opto electronic devices, the possibility to modify the energy band gap as a function of the Qd radius \cite{gapd, proot} can be exploited in order to build light absorbers for photovoltaic applications \cite{abs1,abs2,abs3}. This property can be used to solve the fundamental problem existing with silicon photovoltaic allowing to increase the photo-response of the device in a wide range of the solar spectrum. The spatial confinement in Qds also helps to overcome the Shockley-Queisser limit \cite{shockley, Henry}, i.e. the maximum theoretical efficiency of opto-electronic devices. Besides, discrete energy states appear inside the wide band gap of the insulator matrix, making possible tunable band gap devices \cite{tuneldevices}.\\ 
Nowadays, to increase the conversion efficiency of silicon photovoltaics tandem cells are used \cite{Nozik,Jiang}, i.e., a stack of absorber layers with different bandgaps cover a wider range of the solar spectrum than a single bandgap absorber layer. These structures conform the third generation of solar cells that can increase the energy conversion efficiency limit up to $60\%$ \cite{effi}. Therefore, a whole new generation devices are based on arrays of Qds that form the stack layers, thus, is needed a detailed explanation of the opto electronic response of these arrays.\\
The movement of the electrons under illumination conditions in the Qds are governed by three basic mechanisms: i) promotion to an excited state assisted by a photon (photon absorption), ii) relaxation transitions from excited states to ground state (stimulated photon emission) and iii) tunneling processes to the neighbor Qds or electrodes, since the phonon scattering is suppressed by the phonon bottleneck effect \cite{phonon}. Therefore, in order to generate a net photo-current, the photo-generated electron has to be extracted before recombining, thus, the charge transport from the location of photo-generation to the leads becomes a crucial point that governs the response of the system. Indeed, to obtain a good optoelectronic device not only the photo-generation has to be effective, also the charge transport extraction mechanisms must be efficient to enable electronic current flow. These transport mechanisms depend on tunneling processes thus being strongly dependent on the insulator material and also on the geometrical arrangement of the Qds. \\
A compact model to explain the elastic electrical transport in the self-consistent field regime \cite{suplibro} in these kind of systems was reported by the authors previously \cite{illeraEPL,illeraJAP,illeraArxiv}. It is based in the Transfer Hamiltonian formalism \cite{payne} in order to describe the currents between the different parts of the system. Assuming that the Qds are independent, the non equilibrium distribution function can be obtained solving a rate equation \cite{gur}. Moreover, the effect of the trapped charge in the Qd has been taken into account solving the transport equations and the Poisson equation self consistently, within the self-consistent field regime, explaining a first order the Coulomb blockade effects \cite{suplibro}.\\
The aim of this work is to extend the previous developed transport formalism \cite{illeraEPL} and incorporate the optical processes in order to create a complete model to describe and simulate opto-electronic devices from a small number of basic material parameters. The rate equations have been modified in order to reflect the stimulated generation/recombination carrier processes and the transition rates have been derived from the classical electron light interaction taking into account all the possible transitions due to the optical processes. We present here a complete and compact methodology that can derive the electronic and optical properties and the response of a device based on Qds embedded in an insulator matrix placed between two electrodes from basic material parameters.\\
Some cases based on Si Qd embedded in $SiO_2$ matrix have been studied in detail, the single Qd in two different scenarios, symmetric and asymmetric configuration. Different rates between the optical and current terms have been studied obtaining the key features in order to obtain a net photo-current without external bias voltage applied. Once this case has been studied, two more cases were analyzed, the parallel and serial case. Both cases were studied under external illumination and applied voltage. Finally, in order to validate the presented methodology it was compared to experimental results extracted from Prins et al. \cite{Prins} obtaining successfully agreement between the simulations and the experiments.  

\section{Theoretical model}
The general framework has been presented in previous works \cite{illeraEPL, illeraJAP}. It describes transport between N Qds embedded in a dielectric matrix coupled to two electrodes (L lead and R lead) as a network of multi-tunnel-junctions. It is based on the non coherent rate equations \cite{aver,gur} and the Transfer Hamiltonian approach \cite{payne}. In that framework, each Qd is treated independently and the non-equilibrium distribution function of each Qd can be obtained. This model provides intuitive results and it has successfully been validated to other methodologies like Non Equilibrium Green Function Formalism (NEGFF) with positive results \cite{illeraJAP}.\\
From the Transfer Hamiltonian approach, the currents between the different parts of the system are described by the density of states of each part, the different occupation probabilities and the transmission probability trough the insulator barrier. The charge evolution inside each dot can be written and solved in the steady state extracting the non equilibrium distribution function of each Qd.\\
Moreover, the influence of the applied bias voltage has been introduced taking into account the capacitive coupling between the different elements of the system. The effect of the accumulated charge in each dot is also taken into account within the self-consistent field regime.\\
In order to describe the system as realistic as possible all the variables that appear in the model have been expressed as a function of basic material constant. For the transmission coefficients, a WKB expressions have been used in order to describe the tunnel probability under low and high electric fields. The capacity values have been described using expressions between a infinite plane and a sphere and between two spheres to calculate the Qd-lead and Qd-Qd capacities respectively. Therefore, the complete system can be fully determined by the material constants and some geometrical parameters. A complete explanation of the used expressions are presented in Ref.~\cite{illeraArxiv}.\\
The previous model is used to evaluate the electrical response of the system under external bias voltage, our model considers that the electronic structure inside the Qd is not affected by external influences (light interaction or external electrical polarization), therefore, in the previous equation the equilibrium properties of the Qd are used: effective masses, confinement potentials and energy levels. We use it as a base to develop the photo-response of an array of Qds but some modifications are needed. We introduce the stimulated generation/recombination assuming that the optical and transport processes are independent as proposed by \cite{opticalrateeqIEE,opticalrateeqPRB}. Therefore, the rate equations for each energy levels are modified as 
\begin{eqnarray}
\label{rateeq}
\frac{d n^i_{j}}{dt}=\underbrace{\frac{2q^2}{\hbar}|T_{Lj}|^2\rho_L\rho^i_j(f_{Lj}-n^i_{j})}_{\mbox{Left lead contribution}}+\underbrace{\frac{2q^2}{\hbar}|T_{Rj}|^2\rho_R\rho^i_j(f_{Rj}-n^i_{j})}_{\mbox{Right lead contributions}}+\underbrace{\sum_{k,i'\neq i}\frac{2q^2}{\hbar}|T_{jk}^{ii'}|^2\rho^i_j\rho^{i'}_k(n_{k}^{i'}-n_{j}^{i})}_{\mbox{Neighboring Qds contribution}}  \nonumber \\
+\underbrace{\sum_{k}qR_{kj}\frac{\rho^i_k}{\rho^i_j}n_k^i(1-n_j^i)-\sum_{k}qR_{jk}n_j^i(1-n_k^i)}_{\mbox{Light terms}},
\end{eqnarray} 
where the subscript $i$ refers to the $i^{th}$ Qd, $j$ and $k$ refer to the $j^{th}$ and $k^{th}$ energy level of the corresponding Qd. $n^i_{j}$ is the non-equilibrium distribution function of the $j^{th}$ level in the $i^{th}$ Qd. q is the electronic charge and $\hbar$ is the reduced Planck constant. $\rho_L$ and $\rho_R$ are the density of states (DOS) of the leads evaluated at the energy of the energy level $j$, and $\rho_j^i$ is the degeneracy of the $j^{th}$ energy level of the $i^{th}$ Qd. $f_{Lj}$ and $f_{Rj}$ are the distribution functions of the leads that we consider in equilibrium. Therefore, they are described by the equilibrium distribution function, Fermi Dirac, with the corresponding electrochemical potential, $\mu_L-\mu_R=qV$, where V is the applied bias voltage. $|T_{Lj}|$ and $|T_{Rj}|$ are the transmission coefficients between the Qd and the leads. If the $j^{th}$ energy level belongs to the conduction band we will use the barrier for electrons otherwise we will use the barrier for holes. \\
Concerning the neighboring contribution, three kinds of processes are taken into account; 1) the conduction-conduction band transmission, 2) the conduction-valence band transmission and 3) the valence-valence band transmission depending on the energy levels involved in the tunnel process \cite{illeraArxiv}. The neighboring Qd contributions are summed over all the Qds, except for that under consideration.  \\
The last terms correspond to the generation and recombination terms due to the light interaction \cite{opticalrateeqIEE}. These terms involve all the energy levels of the $i^{th}$ Qd as a function of the optical transition values $R_{jk}$. The first light term is the generation rate while the second one is the recombination. The coefficient $R_{jk}$ is the transition probability between the levels $j$ and $k$, it is explained in the next section. As expected, the generation/recombination terms depend on the occupancy of the energy levels involved in the optical transitions \cite{mckelvey,klingshirn,cardona}. \\
Solving Eq.~\ref{rateeq} in the steady state condition, the non-equilibrium distribution function for each energy level of each Qd can be obtained. Therefore, the charge stored in the $i^{th}$ Qd is $N_i=\sum_j2\rho_j^i n_j^i$, where we take into account the spin factor 2. Once the charge has been calculated, the Poisson equation is solved and the self-consistent solution of the local potential and the charge is imposed \cite{Sup2}.   

\subsection{Optical properties}
As we have seen in Eq.~\ref{rateeq} the transport properties depend on the DOS of the Qds which it describes the electrical and optical properties of the Qds. Therefore, a correct description of the electronic structure of the embedded Qd is needed.\\
Following our strategy to keep the model as general as possible, we propose a simplified model to describe the Qds electrical and optical properties based on the effective mass approximation (EMA) \cite{Sup}. The Qd is treated  as a finite spherical potential well \cite{confinament}. The confinement of the carriers is described by the potentials $V_e({\bf{r}})$ and  $V_h({\bf{r}})$ for electron and holes, respectively. The height of the potential well is the difference between the conduction/valence band energy level of the dielectric matrix and the ones that form the Qd. The width of the well is $2R$, where R is the Qd radius. We use EMA to describe the electrons by replacing the free electron mass $m_0$ by an affective one, denoted $m^*$, which takes into account the periodicity and the interaction with the atomic structure. Moreover, the carriers effective masses $m^*$ are assumed to have different values in the Qd and in the dielectric matrix.\\
The total Hamiltonian of the system has three different parts: the electron/hole parts (kinetics and confinement terms) and Coulomb interactions (e-e, h-h and e-h) \cite{excitontheory}. The e-e and h-h interaction are treated within the self-consistent field regime (SCF) \cite{suplibro}. As a first approximation, we neglect the coupling of different bands (e-h interaction). Many authors have studied excitonics effects in Qds described by the infinite/finite quantum confinement in the EMA approximation including the Coulomb repulsion in the strong confinement regime \cite{EMAQds1, EMAQds2}. The bulk Bohr radius exciton is defined as $a_B=0.529 \epsilon m_o (\frac{1}{m_e}+\frac{1}{m_h})$ \cite{Radioexciton} where $\epsilon$  is the dielectric permittivity of the Qd, $m_e$ and $m_h$ are the effective masses for electrons and holes respectively.  When the Bohr radius is larger than the Qd radius the strong confinement regime is achieved \cite{Limitestrongconf} and the e-h interaction can be treated as a perturbation. Therefore, the total Hamiltonian is separated into independent electron and hole parts. In this regime, the electron and hole can be treated as independent particles solving their respective Schrodinger equations, and the Coulomb interaction between them is a perturbation \cite{Vcoulpert}. This interaction adds a new energy contribution that reduces the optical effective gap. Thus, the total energy has a contribution that depends on $1/R^2$, from the quantum confinement, and $1/R$, from the repulsion. Coulomb interaction only shifts to the red the energy of these transitions \cite{redshift}. Thus, the electron hole pair (exciton) is well described by the product of the single-particle electron hole wave function $\psi_{eh}(r_e,r_h)=\phi_e(r_e)\phi_h(r_h)$ where we have omitted the spin terms for clarity. Moreover, the exciton energy is the sum of the single-particle energies.\\
From now on, we write all the calculations for the conduction band, the valence band calculations are straightforward.  The single particle radial Schrodinger equation is written as 
\begin{eqnarray}
 \left \{-\frac{\hbar^2}{2}\vec{\nabla}_r ( \frac{1}{m^*(r)}\vec{\nabla}_r )+\frac{l(l+1)\hbar^2}{2 m^*(r)r^2}+V(r) \right \} R_l(r)=E_l R_l(r).
\end{eqnarray}
Here, $m^*(r)$ is the position-dependent electron's effective mass, $l$ is the angular momentum quantum number, $V(r)$ is the finite confining potential, $E_{l}$ is the electron energy eigenvalues for the quantum number $l$ and $R_{l}(r)$ is the radial wave function. The angular dependence of the wave function, $Y(\theta,\phi)$, is given by the spherical harmonics. The general solution of the wave function is
\begin{equation}
\phi_l(r,\theta,\phi)=A_l j_l(k_0 r)Y_{lm}(\theta,\phi) \Theta (R-r)+B_l h_l^1(k_1 r)Y_{lm}(\theta,\phi) \Theta (r-R),
\end{equation}
where $j_l(k_0 r)$ is the spherical Bessel function ($l^{th}$ order), $k_0=\sqrt{\frac{2 m_{Qd}^*}{\hbar^2} E}$, $ \Theta (R-r)$ is the step function, $R$ is the Qd radius,  $h_l^1(k_1 r)$ is the spherical Hankel function and $k_1=\sqrt{\frac{2 m_{Ox}^*}{\hbar^2} (V_0-E)}$ \cite{sakurai}. $V_0$ is the band offset between the Qd and the surrounding material, and $m_{Qd}^*$ and $m_{Ox}^*$ are the effective masses in the Qd and in the oxide, respectively. $A_l$ and $B_l$ are the normalization constants. The boundary conditions of the wave function and its first derivate require 
\begin{equation}
\label{states}
\frac{j_l(k_0 r)|_R}{\frac{d j_l(k_0 r)}{dr}|_R}=\frac{h_l^1(k_1 r)|_R}{\frac{d h_l^1(k_1 r)}{dr}|_R}.
\end{equation}
From these conditions, the eigenvalues for each $l$ binding states are obtained. We remark that the degeneracy of each binding state is $(2l+1)$, which implies that the number of binding states increases dramatically when the Qd radius increases. Up to here, the internal structure of the Qd is described.\\
In the presence of a photon field, the effective electron/hole photon interaction can be introduced via the minimal coupling in the Hamiltonian \cite{sakurai}. Assuming that the photon can be treated as a perturbation, the interaction can be calculated starting from the Fermi's Golden Rule in the dipole approximation. This is the usual treatment of the light interaction and a formal derivation can be found in many text books \cite{sakurai, cardona,klingshirn}. These transitions rates correspond to interband transitions. In a general form, the transition rate between the l and l' state can be written as  
\begin{equation}
\label{rate}
R_{ll'}=\frac{16 \pi^2 \alpha}{\sqrt{\epsilon_r} V_{Qd} } I(h \nu) |M_{l'l}|^2 \delta(E_{l'}-E_l-h \nu),
\end{equation}  
where $\alpha$ is the fine structure constant, $\epsilon_r$ the Qd dielectric permitivity, $V_{Qd}$ is the Qd volume, $I(h \nu)$ is the flux of the incident light and $\nu$ is the photon frequency. $M_{l'l}$ is the dipole matrix element averaged over all polarizations of the incident light
\begin{equation}
|M_{l'l}|^2=\frac 1 3 \left\{|x_{l'l}|^2+|y_{l'l}|^2+|z_{l'l}|^2 \right\}.
\end{equation}
Where the polarization in the z-direction is
\begin{equation}
z_{l'l}= \int_0^{\infty} R^*_{l'}(r)R_l(r)r^3 dr \int Y^*_{l'm'} Y_{lm} cos\theta d \Omega,
\end{equation}
we have used $z=rcos\theta$. It is straightforward to calculate the matrix elements in the other two directions. The selection rules for such transitions are determined by the angular part of the integral. It can be readily shown that the allowed transitions must fulfill the condition $|l-l'|= \pm1$ and $|m'-m|=0,\pm 1$. In addition, the oscillator strength which is related to the dipole transition can be expressed as $P_{l'l}=\frac{2m}{\hbar^2} \triangle  E_{l'l} |z_{l'l}|^2$, where $\triangle E_{l'l}=E_{l'}-E_l$ refers to the energy difference of the two levels.\\
Concerning to the intraband transitions, the relevant dipole matrix elements are those between the vacumm and the single exciton states. Using the dipole operator \cite{quantumopticals} the matrix element is written as
\begin{equation}
\label{exciton}
|M_{e-h}|^2=|\int dr^3 \phi_e^*(r) \phi_h(r) |^2,
\end{equation}
where $\phi_e$ and $\phi_h$ are the single particle functions (we have omitted all the quantum number associated to each state). From Eq.~\ref{exciton}, we see that this term involves states in different bands, creating pairs of electrons and holes with the same quantum numbers. The transition rate for this process is defined by Eq.~\ref{rate} but using the corresponding matrix element and the exciton energy.\\
From Eq.~\ref{rate}, the absorption coefficient related to all the different transitions, intraband and interband, is written as
\begin{equation}
\label{alpha}
\alpha(h \nu)=\sum_i \frac{16 \pi^2 \alpha}{\sqrt{\epsilon_r} V_{Qd} } h \nu |M_i|^2 \delta(\triangle E_i-h \nu),
\end{equation}
where the subscript $i$ refers to all the processes and $\triangle E_i$ is the energy level difference of the initial and final state. This summation is performed taking into account the different occupation of the initial and final states. To obtain a smooth absorption spectrum, we replace the $\delta$ function in Eq.~\ref{alpha} with a Lorentzian function with a half-width $\Gamma$which is related to the thermal broadening.

\section{Results and discussion}
\begin{figure}[h!]  
\centering{\includegraphics[width=0.5\textwidth]{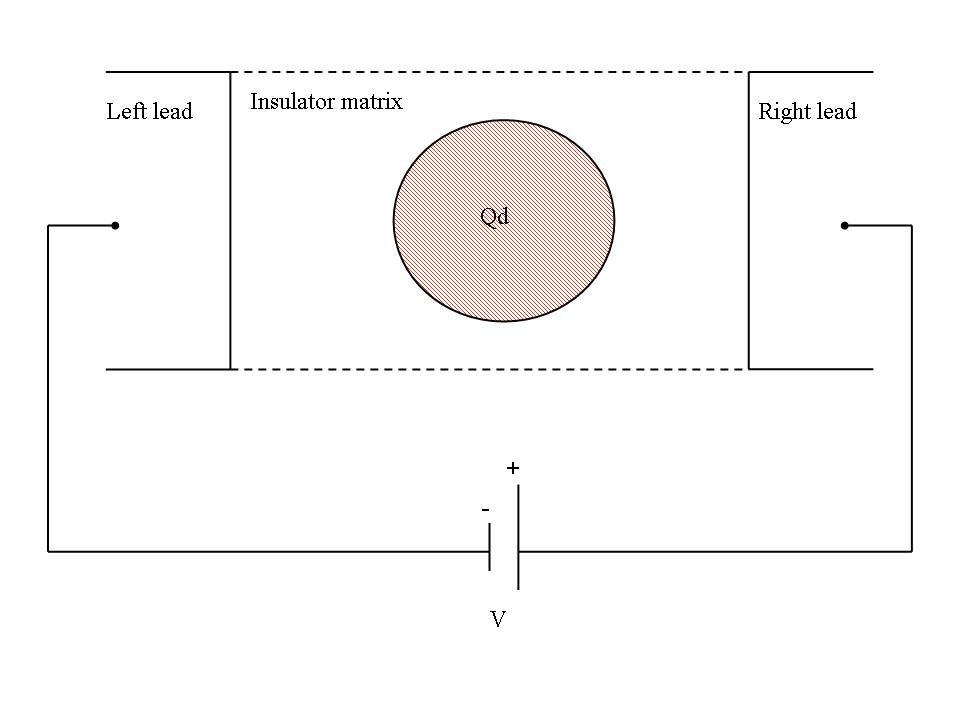}}
\caption{\label{System} Scheme of the basic system. The two leads, left and right, and the central region where the Qd is embedded in the dielectric matrix. An external bias voltage is applied and the Qd can also be illuminated by an external light source.}
\end{figure}

\begin{figure}[h!]  
\centering{\includegraphics[width=0.5\textwidth]{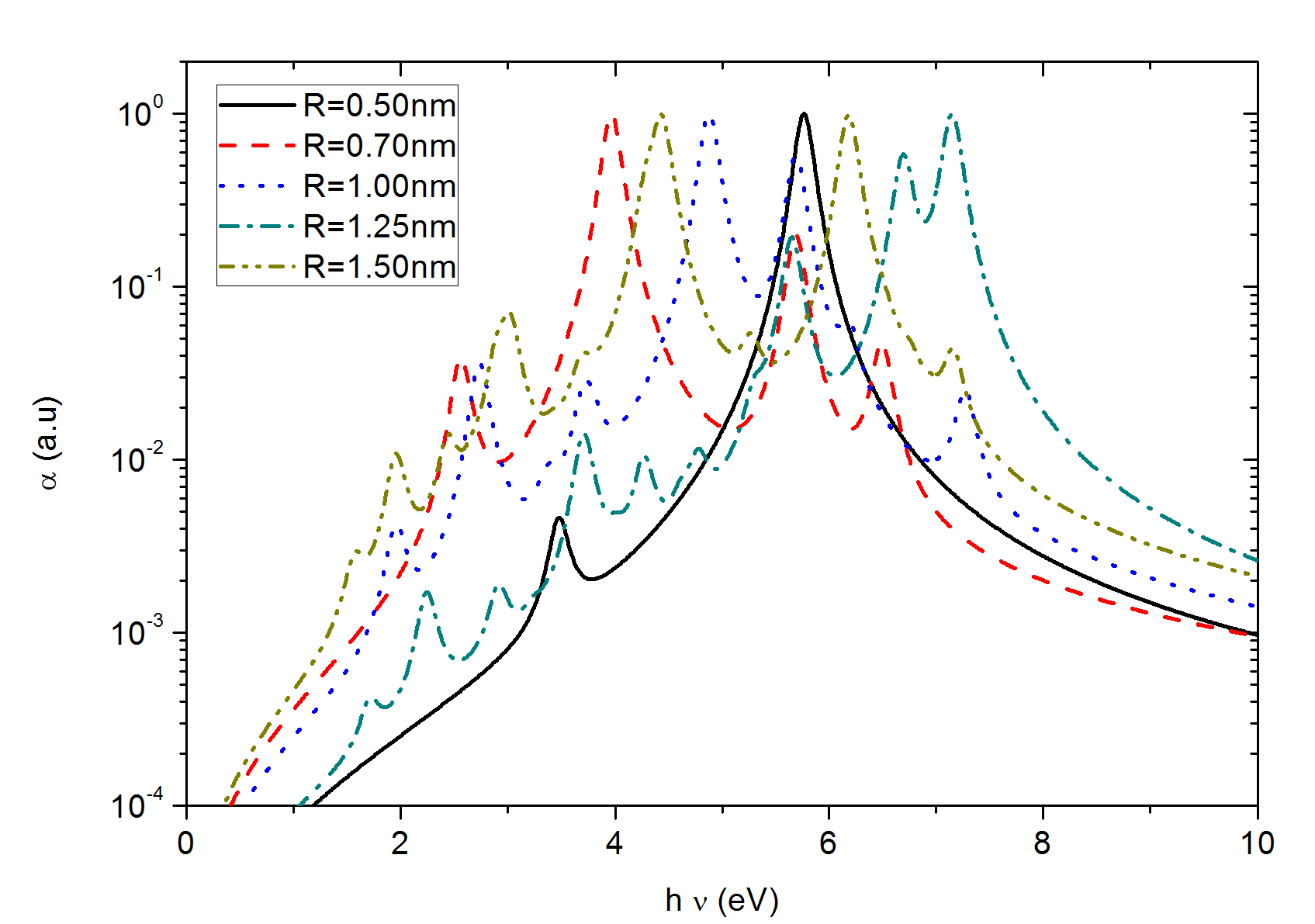}}
\caption{\label{alphar} Simulated absorption spectra in the equilibrium state for different Qd radii.  The summation of all the contributions to the spectra have been done according to the Fermi Dirac distribution function $f(E)=\frac{1}{exp(E-\mu)/kT+1}$, where for simplicity the Fermi level has been placed in the energy origin, $\mu=0$.  The number of resonant peaks increases with the radius and the position of the first peak decrease since the energy level difference decreases.}
\end{figure}

We have done a systematic study of the electrical and optical response of a system composed by two electrodes, left and right, and a central region in which the Qds have been placed. The scheme of the system is shown in Fig.~\ref{System}. The Qds were embedded in an insulator matrix. In our model, the system is described by the geometrical properties of the Qds, radius and relative position respect to the leads or other Qds and the material parameters that describe the insulator and the Qds. In Table I, we show the list of parameters used to describe Si Qds embedded and in $SiO_2$ insulator \cite{illeraArxiv}. The effective masses inside the Si Qd ($m^*_{Qd,CB}$ and $m^*_{Qd,VB}$), in the oxide ($m^*_{ECB}$, $m^*_{EVB}$ and $m^*_{HVB}$), the confinement potentials $\phi_{1,ECB}$ and $\phi_{1,HVB}$ for electrons and holes respectively, the Si bulk gap $E_{gSi}=E_{shift,CB}+E_{shift,VB}$, and the $SiO_2$ dielectric permittivity $\varepsilon_r$. We use T=300K.\\
\begin{table}[h!]
\begin{center}
\begin{tabular}{|c |c| |c| c|}
\hline
$m^*_{ECB}$ ($m_0$) & $0.40$ & $\phi_{1,ECB}$ (eV) & 3.1\\
$m^*_{EVB}$ ($m_0$) & $0.30$ & $\phi_{1,HVB}$ (eV) &-4.5\\
$m^*_{HVB}$ ($m_0$) & $0.32$ & $E_{shift,CB}$ (eV)& 0.6\\
$m^*_{Qd,CB}$ ($m_0$) & $1.08$ &  $E_{shift,VB}$ (eV)& -0.6\\
$m^*_{Qd,VB}$ ($m_0$) & $0.57$ & $\varepsilon_r$ ($\varepsilon_0$) & 3.9\\
\hline
\end{tabular}
\caption{\label{tabla}Parameters used in the simulation in order to describe Si Qds embedded in $SiO_2$ insulator matrix. $m^*_{ECB}$, $m^*_{EVB}$ and $m^*_{HVB}$ refer to the effective masses used to describe the different tunneling processes: conduction-conduction band transmission, conduction-valence band transmission and valence-valence band transmission, respectively. $E_{shift,CB}$ and $E_{shift,VB}$ are the energy shift of the binding states due to the Si bulk band gap.}
\end{center}
\end{table}
The study of the energy states of the Qd has been done assuming the finite spherical potential well using the Si and $SiO_2$ effective masses and the confinement potentials assuming bulk values. The binding states appear from the numeric solution of Eq.~\ref{states}, a detailed description of the $E_{gap}(R)$ and the DOS of Si/SiO2 Qds is presented in Appendix A.1. Since the eigenstates and the wave functions have been obtained; the optical properties of the Qd, the absorption spectra, can be calculated. Using the effective masses described in Table I and the bulk Si permittivity $\epsilon_r=11.7$, the bulk exciton Bohr radius is estimated as $a_B=16.58$ \AA. The absorption spectra gives important information about the electron transitions between different energy levels produced by the interaction with an external photon and its probability. We present, in Fig.~\ref{alphar}, the evolution of the simulated spectra as a function of the Qd radius using Eq.~\ref{alpha} for a z-polarized incident light. When the Qd becomes larger, the number of energy states in the Qd increases obtaining more possible transitions. Moreover, the value of $E_g$ decreases \cite{proot} thus, a redshift of the spectra is observed. For high photon energies the absorption decrease since we only consider the transitions between the discrete energy states of the Qd. Therefore, we neglect the continuum conduction and valence band states and all the possible transitions from/to these bands.\\
Once the absorption spectra is calculated, the photo-current response of the different systems can be obtained. The systems are composed by two electrodes, right and left leads and a central region in which the Qds are placed. For simplicity, we assume an incident monochromatic light with irradiance $I(h \nu)=1Wm^{-2}$. Using the rate equation formalism described previously, the current can be obtained under external bias voltage conditions, illuminated or in both cases. 

\subsection{One single Qd}
\begin{figure}[h!]  
\centering{\includegraphics[width=0.5\textwidth]{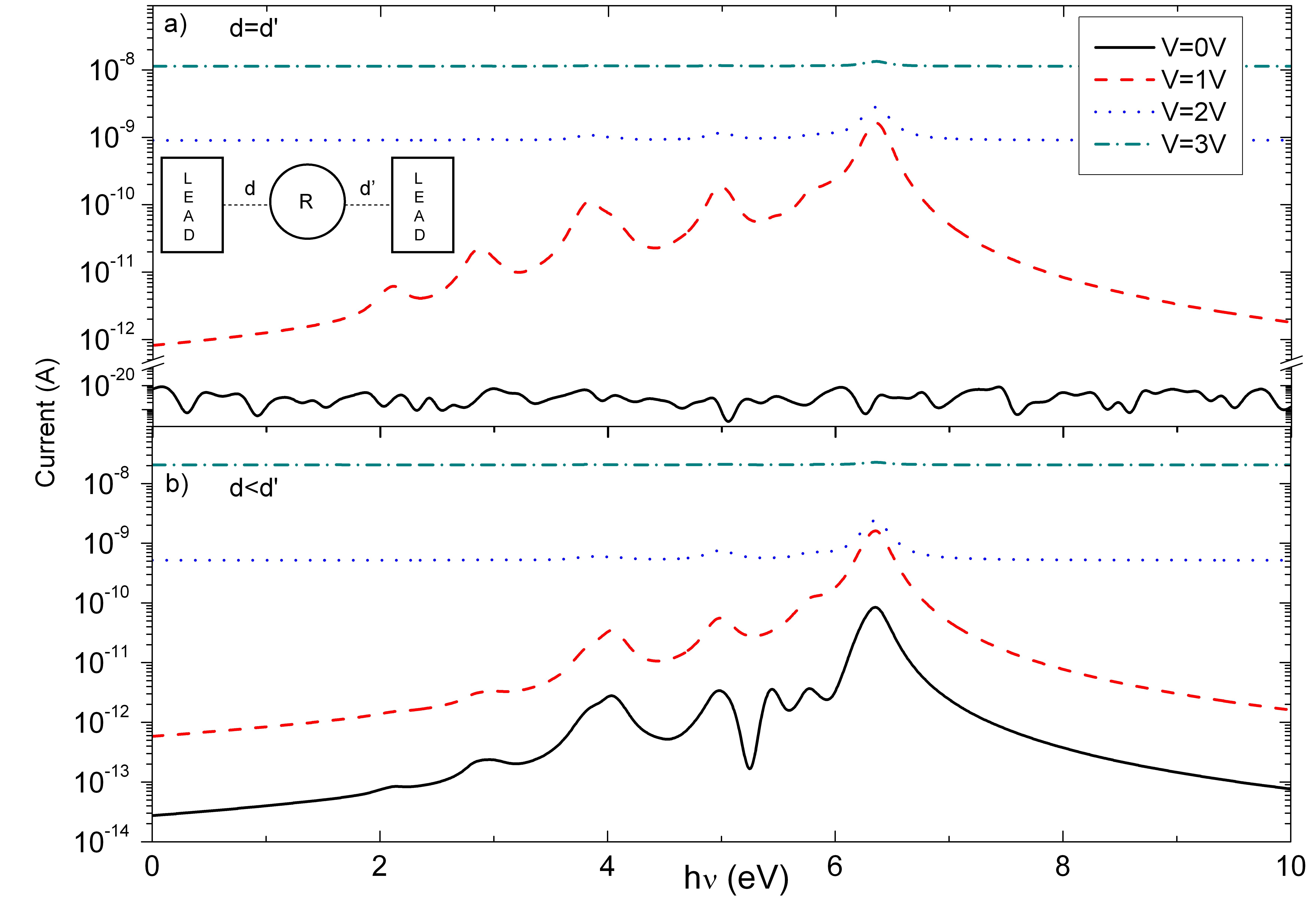}}
\caption{\label{I(hnu)} Photo-current as a function of the energy of the incident light with an external applied bias voltage. In the inset a scheme of the system is presented, a Qd of $R=1.06nm$ is placed between the two electrodes. a) Symmetric system $d=d'=1.78nm$. For $V=0$ case the current is zero while for $V \neq 0$ the symmetry of the system is broken and net current appears. b) Asymmetric system $d=1.47nm$ and $d'=2.09nm$. A net current is obtained even at $V=0$. The current peaks refers the position of the maximum optical transition probabilities.}
\end{figure}
The first system under study is a single Si Qd under illumination connected to two electrodes with a constant external bias voltage applied. From Eq.~\ref{rateeq} many scenarios appear as a function of the value of the optical and transport terms. We study the particular case of $V=0$, if the transport terms are grater than the optical ones the non-equilibrium distribution function of the energy levels will follow the Fermi Dirac distribution function since the optical terms are a small perturbation. When the optical term increases, an equal generation of electron and holes appears. Both carriers tend to diffuse to the electrodes creating currents and, depending on the probability of these transitions, the Qd could be charged. Therefore, depending on the relation between the the transmission and transition probabilities the low and high injection level can be achieved.\\
When an external bias voltage is applied the distribution functions of the two leads differ and a net current appears. Overlapped to this, the optical processes appear filling/emptying levels and adding new conductive channels to the transport enhancing the total current.\\  
The results of the single Qd are shown in Fig.~\ref{I(hnu)}. We present two scenarios: a) the Qd symmetrically connected to the leads and (b) in asymmetric configuration. An interesting result appears for V=0. In the symmetrically coupled system (a), the current is zero since the incoming hole currents for each side equals the outgoing electron currents. Therefore, the net current is zero because the electron and hole currents compensate each other. This result derives intuitively from the rate equation, as it can be viewed as a sum of intensities in each energy level equals to zero. Therefore, in order to generate a net photo-current the symmetry of the Qd respect to the leads must be broken, hence, different coupling to the leads are needed as show in the asymmetric case.\\
When an external bias voltage is applied, the transmissions coefficient between the Qd and the two leads change thus the system becomes asymmetric and a net current appears. The current peaks are related to the maximum transition probabilities for an incident photon reflecting the absorption spectra. When the voltage increases the current tends to be independent of $h\nu$. This effect is the result of competition between two processes, the pure light current term and the external bias voltage term. For small voltages, the optical terms dominate and the optical transition peaks are observed, but when the voltage increases the tunneling currents become the most important terms and the current appears as a photon energy independent.\\
\begin{figure}[h!]  
\centering{\includegraphics[width=0.5\textwidth]{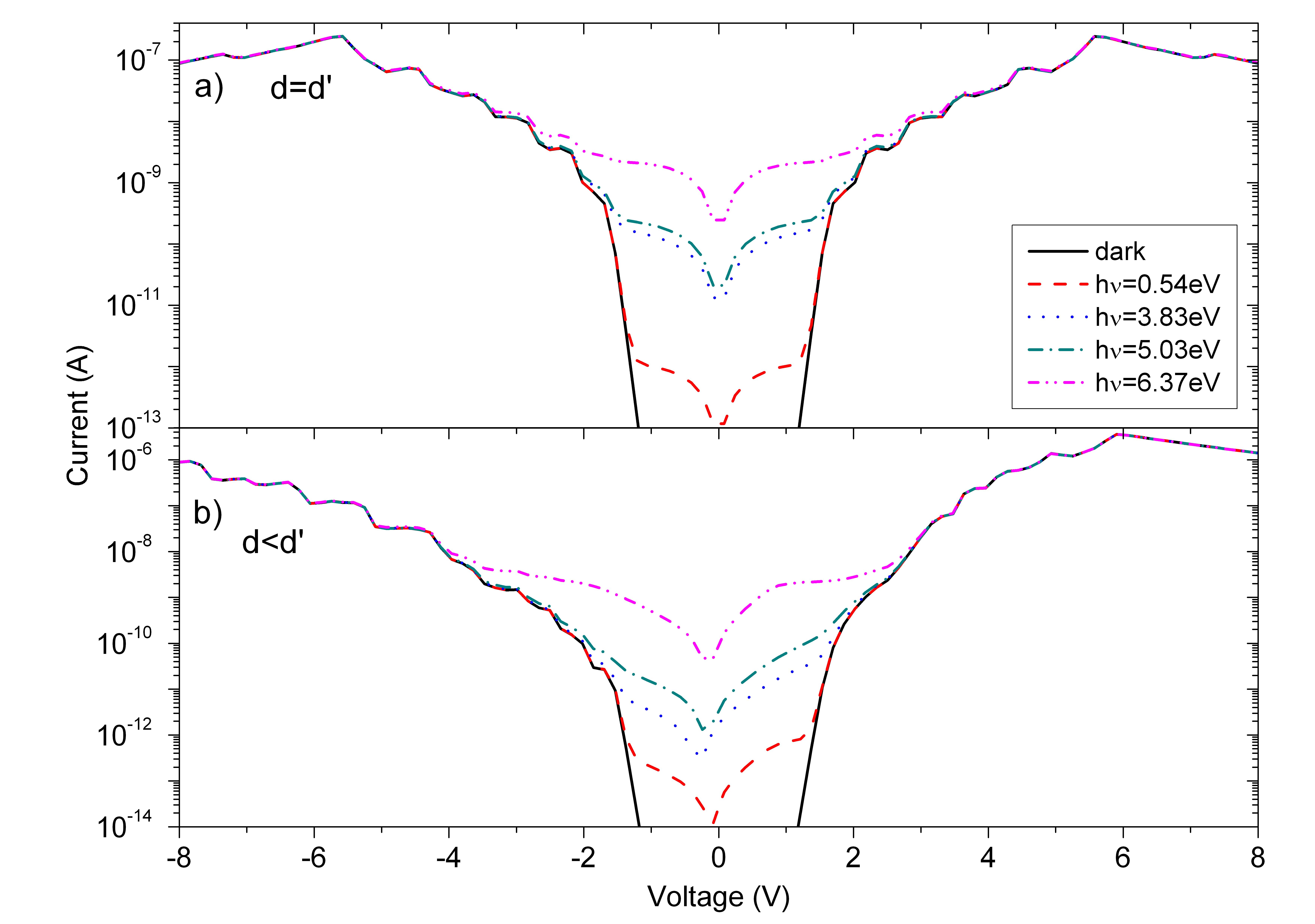}}
\caption{\label{I(V)+hnu}a) The total I(V) curve (in absolute value) for the symmetric system in dark case and under different illumination conditions (incident photon energies). b) The total I(V) curve (in absolute value) for the asymmetric system in dark case and under different illumination conditions (incident photon energies).  }
\end{figure}
In the same way, in Fig.~\ref{I(V)+hnu} we show the obtained current voltage curve I(V) under external illumination for the same previous system as a function of the energy of the incident photon. In  Fig.~\ref{I(V)+hnu} a-b) we show the current for the symmetric and asymmetric structure respectively. As it was demonstrated in previous works \cite{illeraEPL}, the symmetry of the dark I(V) curves depends on the symmetry of the coupling to the leads, and current plateaus appear when conduction channels are opened \cite{Sup}. For the illuminated case, the main differences appear for small voltages. The I(V) curve looses the step like behavior since the occupancy of the energy levels are not sequential with the applied bias voltage due to optical transitions. Moreover, the optical transitions tend to fill the conduction states while the valence states become emptied increasing the electron/hole currents. As a consequence, more conduction channels are opened and the total current is bigger than in the dark case. \\
The next step is the study of a system composed by two Qd. We study two configurations that give us the main properties of bigger systems. In previous works \cite{illeraJAP}, the parallel and serial configurations revealed the basic mechanisms behind the transport in large Qd arrays, therefore, these systems are the best starting point in order to extend the methodology to larger systems.

\subsection{Parallel case}
\begin{figure}[h!]  
\centering{\includegraphics[width=0.5\textwidth]{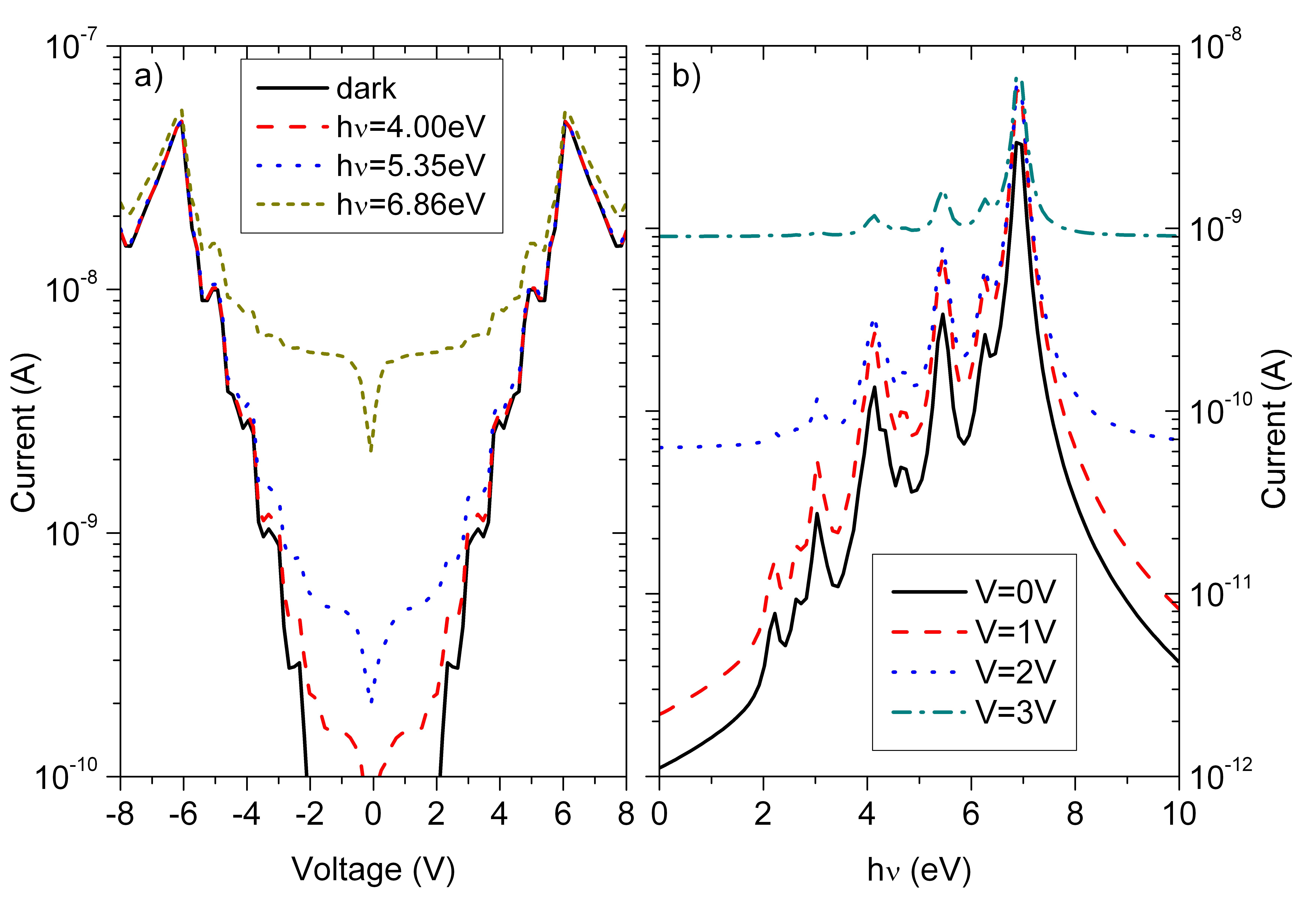}}
\caption{\label{paralelo}The radii Qds are $R_1=1.0nm$ and $R_2=0.8nm$ respectively. The distances between the first Qd ($Qd_1$) and the leads are $d_{L1}=2.5nm$ and for the second one ($Qd_2$) $d_{L2}=2.5nm$. The distance between the Qds is $d_{12}=3nm$. All the distances are measured from the center of the Qds. a) The total I(V) curve (in absolute value) for the parallel system in dark case and under different illumination conditions (incident photon energies). b) Photo-current as a function of the photon incident energy for different applied bias voltage.}
\end{figure}
We start with the case of two Qd in a parallel configuration. We use two Qd of different radius in order to obtain different optical transitions  in each Qd for a given incident photon.  Now, the rate equations for each energy level of the Qds have three contributions; the leads, the neighbor Qd and the optical contributions. In Fig.~\ref{paralelo}a) we show the I(V) curve under different incident photon energy. These photons energies are related to the maximum absorption peaks, for $R=0.8nm$ and $R=1.0nm$ Qds. In a previous study, we have demonstrated that in parallel configuration the total dark current trough the system is the sum of the individual Qd terms \cite{illeraJAP}. At low voltages, the optical terms dominate in the rate equation and the obtained I(V) curve differs from the dark curve. As shown before (for one single Qd), this trend is a competition between the optical transition probabilities and the transmission probabilities. For low voltages, the barriers are not transparent enough. Therefore, electron/holes are photo-generated in the Qd changing the distribution function of the Qd creating net fluxes incoming/outgoing from the Qd to the leads. The electrons tend to move from the Qd to leads and the holes follow the opposite direction, obtaining a net current if the currents are different. When the voltage increases, the barriers are bended and the tunneling probability increases as well. Then, the optical term becomes smaller than the electrical terms and the dark trend is recovered. When the energy of the incident photon increases, the optical transitions involve the higher energy levels (valence and conduction states for electron and holes respectively) which have the maximum transmission probabilities increasing the current.\\
Moreover, in this case, the electrons can also move to the other Qd. Thus, a current between Qds appears and breaks the symmetry respect to the leads obtaining a net current for the case V=0V (Fig.~\ref{paralelo}b). As in the previous case, when the polarization voltage increases the photo-current peaks tend to disappear. More peaks appear in the current term that take into account the photo-current generated  in each Qd since the two Qds have different radii and the optical transitions occur at different photon energy.     

\subsection{Serial configuration}
\begin{figure}[h!]  
\centering{\includegraphics[width=0.5\textwidth]{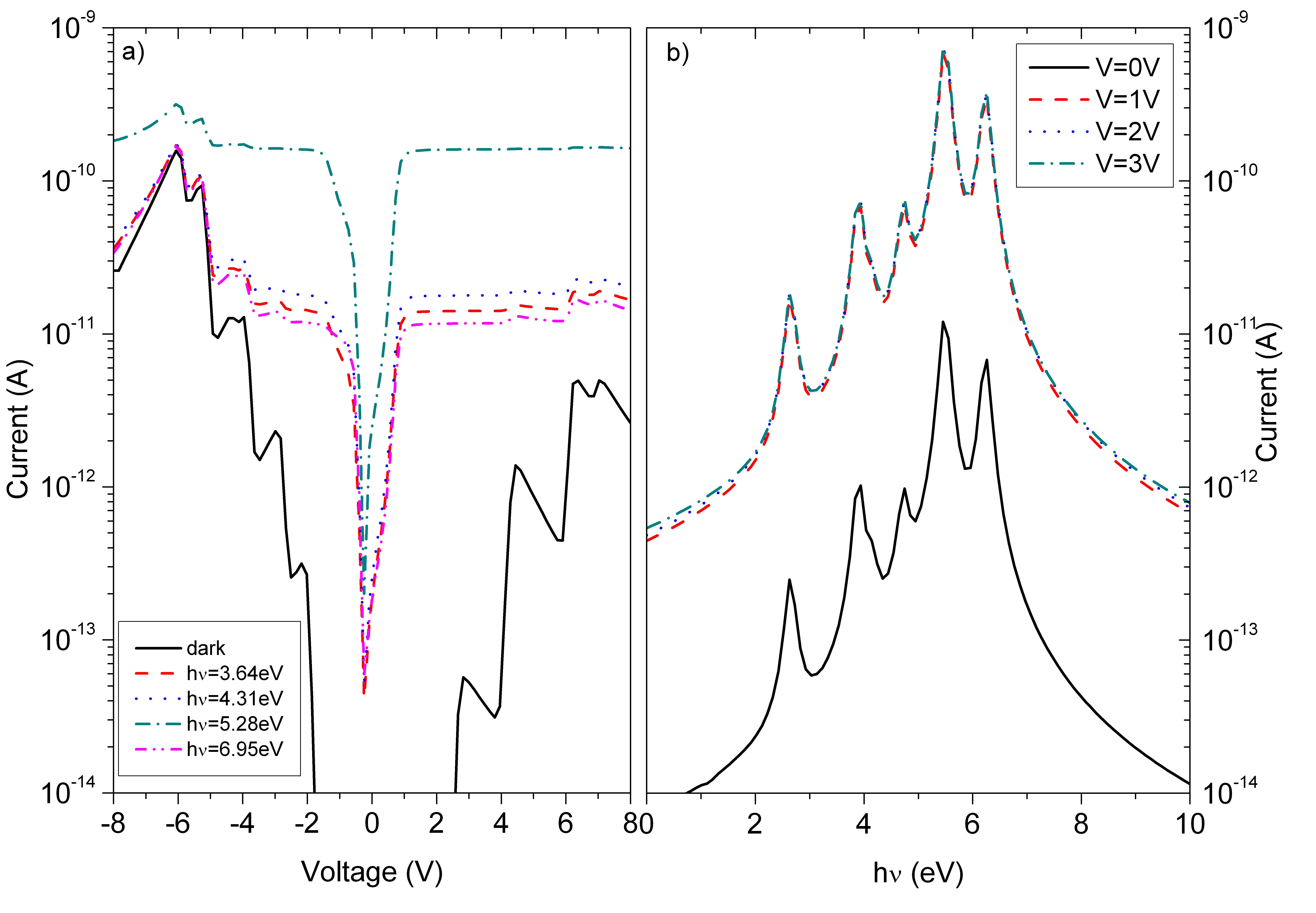}}
\caption{\label{serie}The radii of the Qds are $R_1=0.3nm$ and $R_2=0.8nm$. The first Qd is placed at $d_{L1}=2.5nm$ and $d_{R1}=5.1nm$ and the second Qd is placed at $d_{L2}=4.8nm$ and $d_{R2}=2.8nm$ from the left and right leads respectively. The distance between both Qds is $d_{12}=2.3nm$. All the distances are measured from the center of the Qds. a) The total I(V) curve (in absolute value) for the serial system in dark case and under different illumination conditions (incident photon energies). b) Photo-current as a function of the photon incident energy for different applied bias voltage.}
\end{figure}
The third type of arrangement is two Qd in a serial configuration. In previous works, this kind of arrangement demonstrates the possibility of filtering the current trough the position of the energy levels \cite{RevModPhys, illeraJAP}. Since overlapping between the energy levels of the Qds is necessary, only a few conduction channels are opened and the current is strongly dependent on the electrostatic coupling of the Qds obtaining NDR in the current-voltage curve \cite{RevModPhys2, illeraEPL}. We proceed in a similar way as in the previous case, the electrical response under an external bias voltage and under illumination with different photon energy. In this configuration, each Qd is connected to one electrode (left and right) and the neighbor Qd.\\
First of all, we present the obtained dark I(V) and illuminated curve in Fig.~\ref{serie}. In the dark case, the NDR and current resonant peaks are obtained. The explanation related to these effects can be found elsewhere \cite{RevModPhys2}. When the light is applied, the current increases, since the occupation of the higher energy levels increase, making favorable the tunneling processes. For certain photon energies, the current saturates in a voltage region as the transitions involves the higher/lowest energy states of the conduction/valence band. Overlapped to these optical transitions, we have the electronic transport due to the applied voltage. In order to obtain electronic transport between the Qds, the condition of the overlapping of the energy levels has to be fulfilled. Therefore the NDR and the current peaks still remain in the I(V) curve.\\ 
In addition, we present the photo-current generated as a function of the energy of the incident photon in Fig.~\ref{serie}b). The values of the photon energy have been chosen in order to maximize the transitions probabilities in each Qd obtaining different photo generation rates in each dot. The cases under external bias voltages are also presented. We obtain a similar trend as in the previous cases, the current reflects the absorption spectra of the systems obtaining current peaks when the transitions probabilities are maximal in the cases when the photon energy equals the difference between the energy levels involved in the transition. As a difference as in the previous cases, the current saturates when the voltage is increased as a consequence of the no overlapping between the energy levels of the Qds. For these reason, we study this kind of system; a small Qd (with few energy levels) connected in series with a bigger one (with large number of energy levels). Therefore, the small Qd dominates the behavior of the photo-current since it controls the number of conduction channels. In this configuration, the photo-current retains the current peaks when the external voltage increases since the small Qd acts as a current filter.

\subsection{Comparison with experiments}
To conclude, we present the comparison between the presented theoretical model and a real system. The experimental results have been taken from Prins et al. \cite{Prins}. They fabricated a system based in two electrodes separated by 4nm. A single layer of PbSe Qd of 2nm in radius was deposited on top of the electrodes. Therefore, they obtained a system of Qds placed in parallel configuration. This kind of structure is the same as we have studied in the previous section.\\
In order to simulate this new system, we have changed the material parameters to describe the PbSe Qds. PbSe is narrow band gap semiconductor with large Bohr radius ($a_B=46nm$ \cite{PbSeexciton}) and small effective masses.The value of the effective masses and confinement potentials were taken from Pellegrini et al.\cite{barrerasPbSe}. We used $m^*_{Qd,CB}=0.07$ and $m^*_{Qd,VB}=0.06$ for electron and hole effective masses respectively. The confinement potentials were $\phi_{1,ECB}=1.61eV$ and $\phi_{1,HVB}=1.61eV$. The obtained $E_{gap}(R)$ is presented in Appendix A.2 The dielectric constant value was $\varepsilon_r=23 \varepsilon_0$ \cite{epsilonPbSe}. As an example, we created a system composed by six Qds placed in parallel configuration. With this small system, the main trends and the physics behind the device can be studied in a simple form. The Qd radius is generated randomly using a normal distribution with mean radius $<R>=2$nm and $\sigma=0.1$nm. The Qds were placed in parallel configuration but the distance between them and the leads was chosen as a free parameter breaking the symmetry of the system.\\
\begin{figure}[h!]  
\centering{\includegraphics[width=0.5\textwidth]{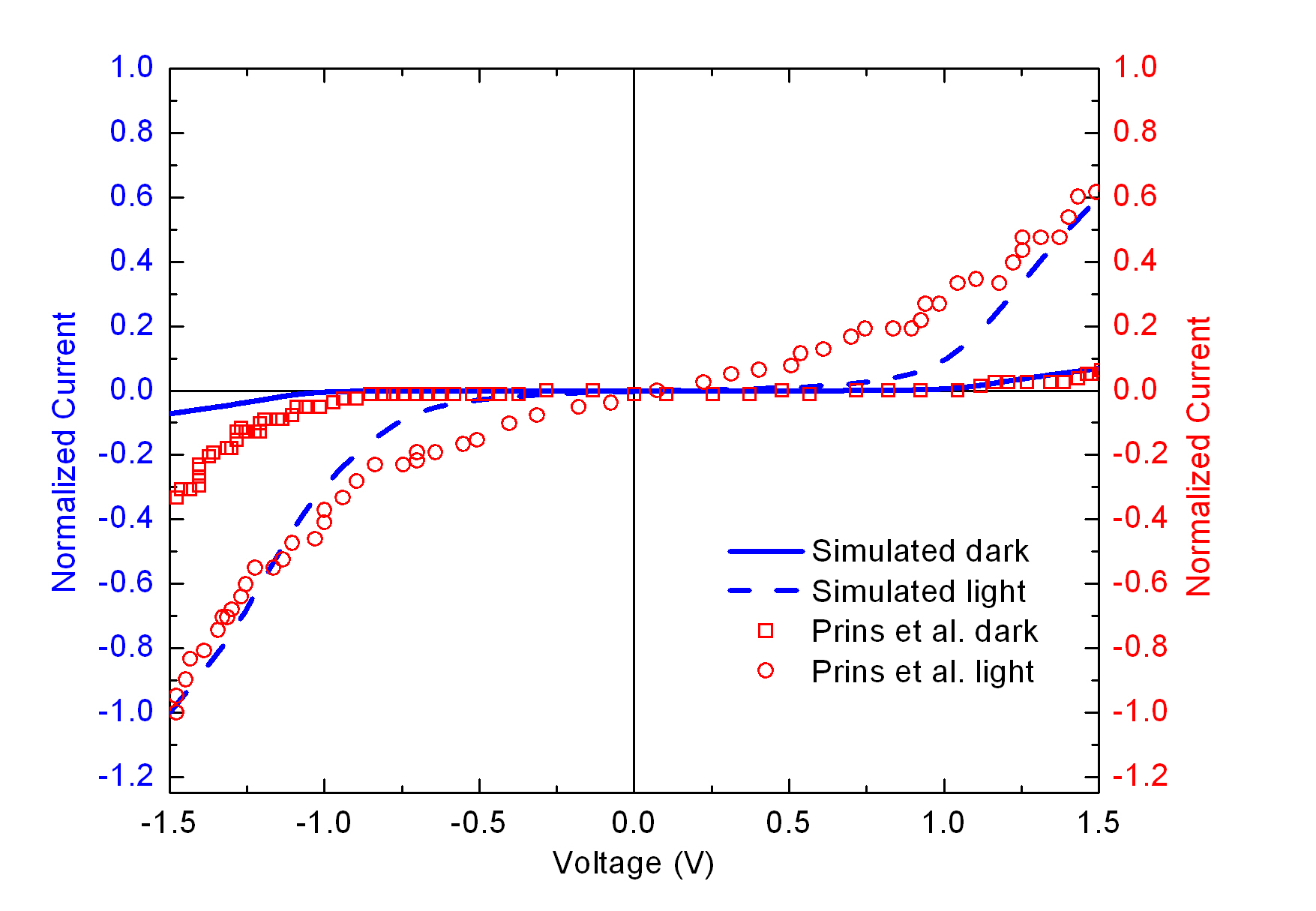}}
\caption{\label{comparacion1} Simulated I(V) curve in dark (blue line) and light (dash blue line) conditions respectively. Measured I(V) points extracted from Prins et al. \cite{Prins} in dark (red open squares) and light (red open circles). The system was illuminated with light of $\lambda=532nm$ and irradiance $I=0.16Wcm^{-2}$.}
\end{figure}
In Fig.~\ref{comparacion1} we present the comparison between the experimental and theoretical I(V) curves in dark and light conditions. The simulated curves follow the trends of the experimental ones in both cases, in dark and with light. The simulated curve still retains the step behavior but when the system becomes bigger and there is a random size distribution of Qds the current contributions of each Qd is summed and the plateaus disappear. Moreover, the theoretical curve reflects the asymmetry in the I(V).\\  
\begin{figure}[h!]  
\centering{\includegraphics[width=0.5\textwidth]{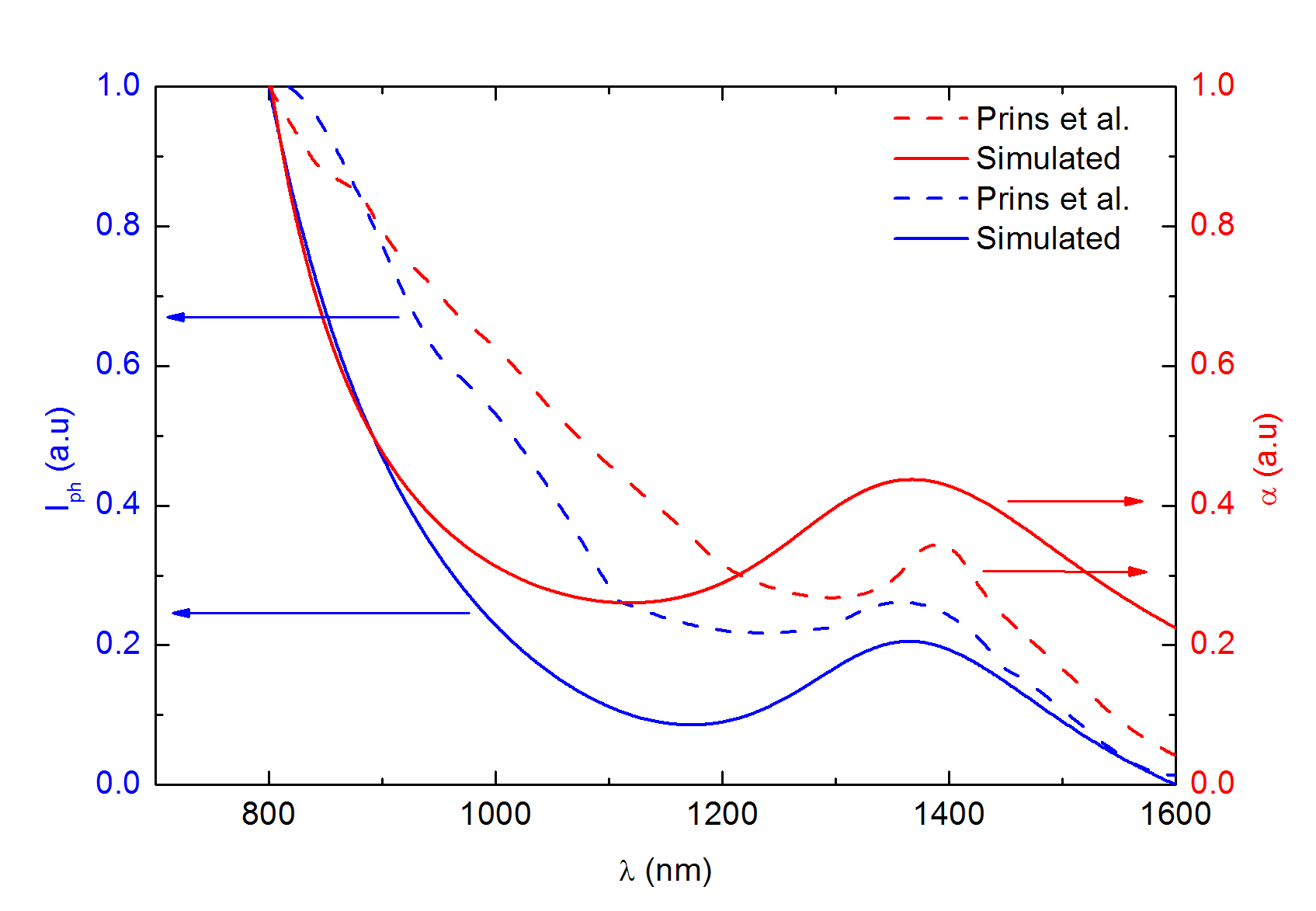}}
\caption{\label{comparacion2} Simulated normalized photo-current curve (blue line) and experimental curve taken from Prins et al. \cite{Prins} (blue open circles) as a function of the wave length of the incident light. The value of the applied bias voltage is V=750mV and the irradiance $I=0.16Wcm^{-2}$.  Simulated normalized absorption coefficient (red line) of the three Qds and experimental absorption (red dashed line).  }
\end{figure}  
We also compute the normalized photo current in Fig.~\ref{comparacion2} and its comparison to Prins et al. Besides, we represent the total simulated absorption coefficient of the system. Concerning the absorption coefficient the simulated reproduces successfully the experimental measurements trend. The simulated spectra reflects the first optical intraband transition around $\lambda=1372nm$ that corresponds to a photon energy $h\nu=0.9048eV$ which is the value of the obtained energy band gap  for Qd radius around $2nm$ (see Fig.~\ref{EgPbSe(R)}). When the photon energy increases, the spectra also increases since there are more available transitions. Therefore, we have reproduced approximately the optic and electronic structure of the Qds which are used as an input for the developed transport model.\\
As we have discussed in the parallel case, the photo current is directly related to the absorption spectra.\\ 
The main goal of the presented transport methodology is that it can be improved using density of states \cite{Nuria} or absorption coefficients \cite{Roberto} derived from powerful techniques as DFT  making suitable a detailed description of the Qds.

\section{Conclusions}
The next generation of optoelectronic devices based on the singular properties of the Qds embedded in an insulator matrix poses new requirements on theoretical models in order to describe their response. We propose a theoretical framework based on the non-coherent rate equations combined with the transfer Hamiltonian approach to explain the transport properties of these kinds of systems in the self-consistent field regime. Within this approach, the rate equations were modified to take into account the optical processes, stimulated generation and recombination. Moreover, the whole methodology presented here is only dependent on basic material parameters and the geometry of the device.\\
This approach has been used to study the electrical response of three systems: one single Qd, two Qds in parallel configuration and two Qd in serial configuration. An important result emerges from the geometrical disposition of the Qds respect to the leads, obtaining that the generated photo-current is zero if the Qd is symmetrically coupled to the leads for the V=0 case.\\
Moreover, the methodology has been used to compare with experimental results. The simulated I(V) curves in dark and under illumination conditions can reproduce in a general trend the experimental observations, making possible the use of this model as a guideline to design new opto-electronic devices. 

\section*{Acknowledgments}
S. Illera is supported by the FI programme of the Generalitat de Catalunya. A. Cirera acknowledges support from ICREA academia program. The authors thankfully acknowledge the computer resources, technical expertise and assistance provided by the Barcelona Supercomputing Center - Centro Nacional de Supercomputaci\'on.

\section*{Appendix}
In this section, we present the obtained results for the value of $E_{gap}$ as a function of the Qd size under the EMA approximation for a finite spherical potential well using the parameters described before for both Qd, $Si/SiO_2$ and PbSe. Moreover, we estimate the binding energy of the ground-state exciton treating the Coulomb interaction as a perturbation as it is usually done in the strong confinement regime. 

\subsection*{$Si/SiO_2$ Qds}
\begin{figure}[h!]  
\centering{\includegraphics[width=0.5\textwidth]{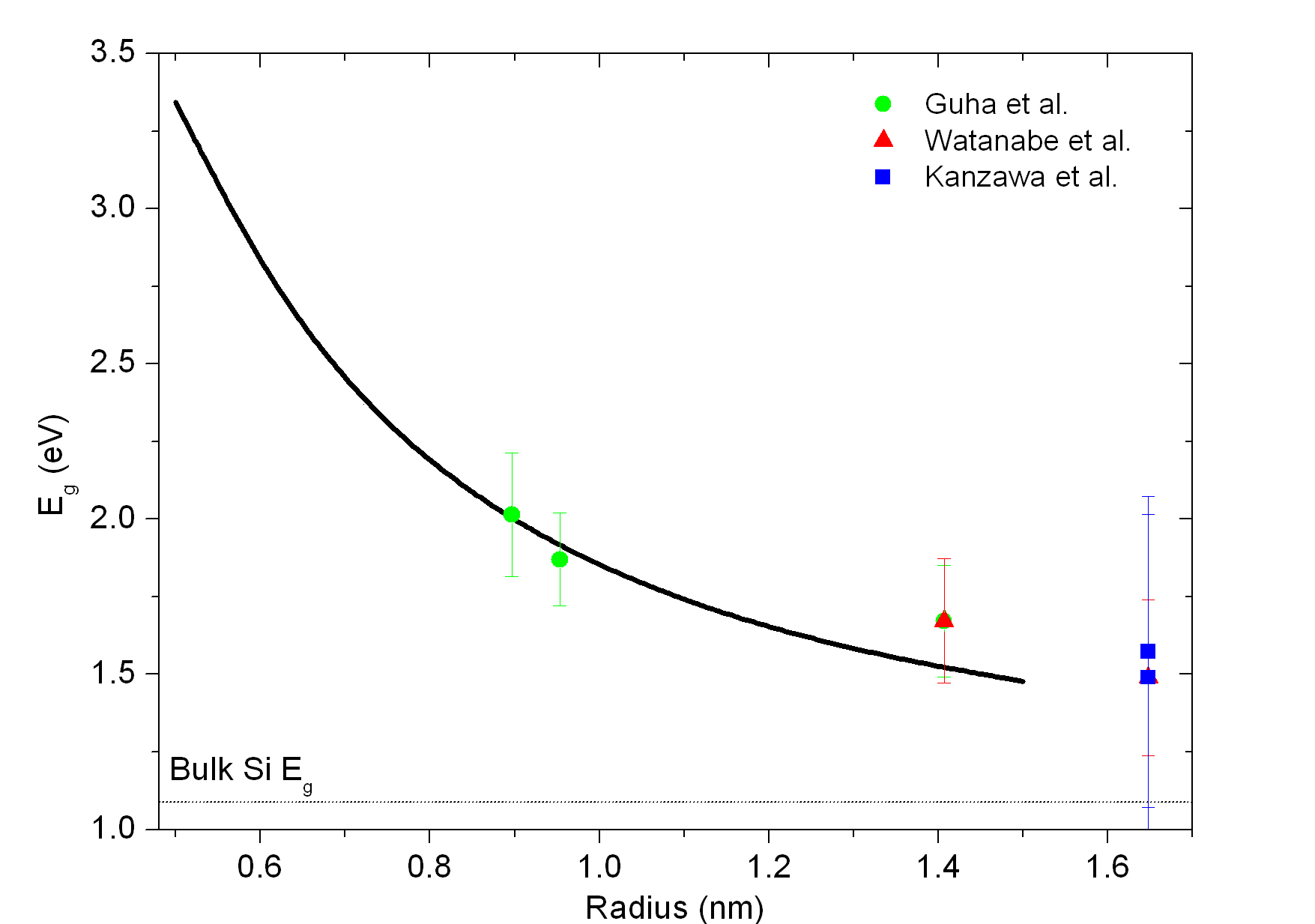}}
\caption{\label{EgSi(R)} Dependence of the obtained $E_{gap}$ as a function of the radius of the Qd without taking into account the exciton shift (solid line). For comparison, experimental data obtained from photoluminescence are presented \cite{Guha,Watanabe,Kanzawa}}
\end{figure}

Here, a comparison between the obtained $E_{gap}$ for the $Si/SiO_2$ Qds using the parameters of Table I and experimental data is presented as a function of the Qd radius. The results are presented in Fig.~\ref{EgSi(R)}.
\\
In Fig.~\ref{EgSi(R)} the agreement between the used EMA approximation and experimental results is presented. An estimation of the energy shift due to the electron-hole Coulomb interaction $E_{e-h}$ can be obtained using the approximation presented in \cite{Coulintegral} for the interaction potential. The energy shift is written as
\begin{equation}
E_{e-h} \approx -\frac{2 q^2}{4 \pi \epsilon_0 \epsilon_r} \int_0^R r_h R_l^2(r_h)dr_h \int_0^{r_h} r_e^2 R_l^2 (r_e)dr_e,
\end{equation}   
where q is the electron charge and $\epsilon_r$ is the dielectric permittivity of the Qd material. For a $Si$ Qd of $R=1nm$ embedded in a $SiO_2$ the estimation of the interaction energy is $E_{e-h}\approx -0.138  eV$ for the first allowed transition, s-hole state $E_{|1sh>}=1.1507eV$ and p-electron state $E_{|1pe>}=1.2243eV$  . This effect decreases the optical band gap but, as can be viewed from Fig.~\ref{EgSi(R)} the effective masses used to describe the Qd agree with experimental theory for small radius ($R<a_B$) reflecting that this effect is included in the effective mass value.    

\subsection*{PbSe Qds}
\begin{figure}[h!]  
\centering{\includegraphics[width=0.5\textwidth]{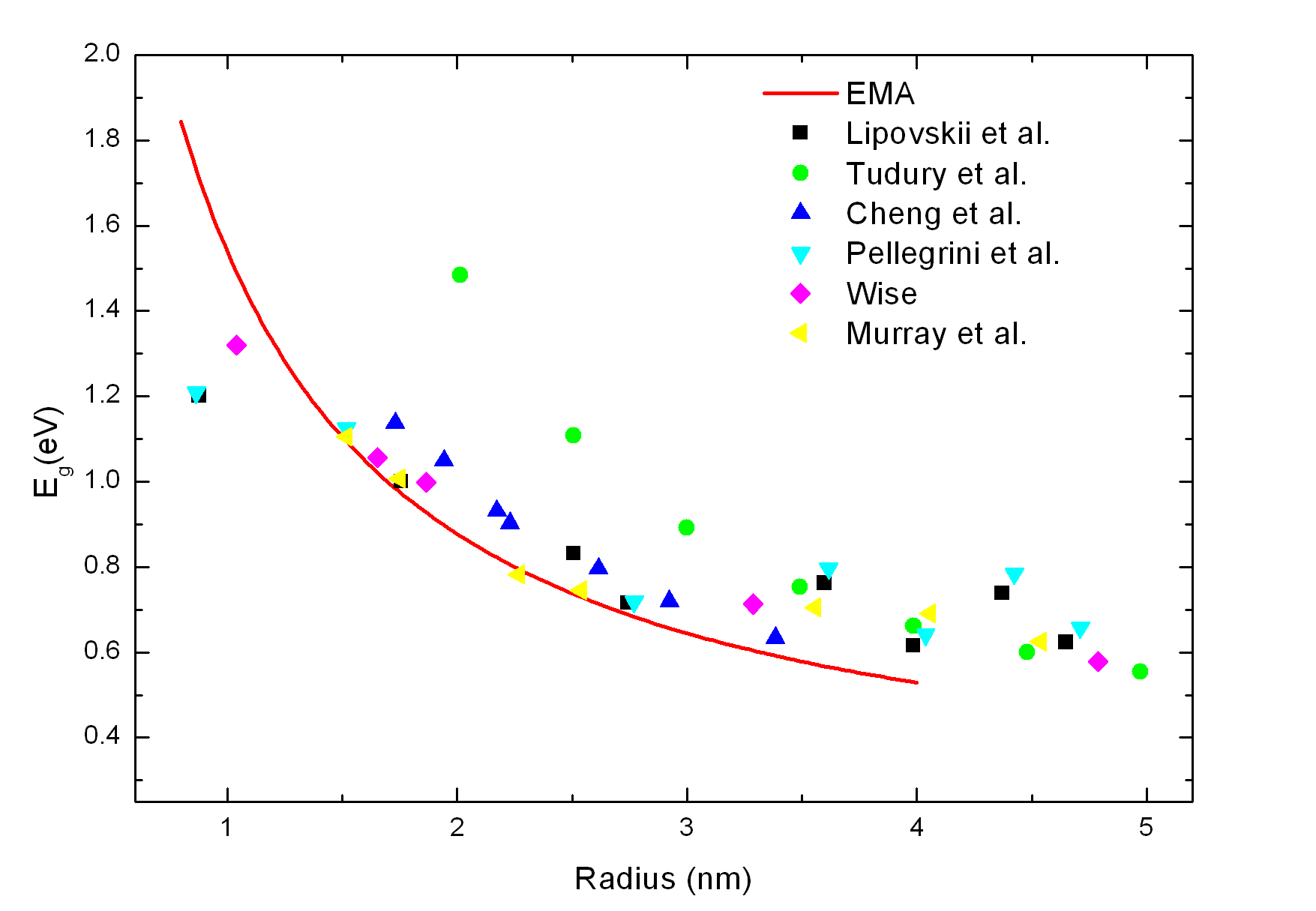}}
\caption{\label{EgPbSe(R)} Dependence of the obtained $E_{gap}$ as a function of the radius of the Qd without taking into account the exciton shift (solid line). For comparison, experimental data \cite{Wise,ChristopherB,barrerasPbSe} and results from different theoretical approaches are also presented \cite{Lipovskii,Tudury,Cheng}.}
\end{figure}

The PbSe Qd is treated as a finite spherical potential well under the EMA approximation neglecting the Coulomb interaction. It has been widely studied under several models as EMA \cite{Brus}, {\bf{K$\cdot$ p}} Hamiltonian \cite{Kpmodel}, finite barrier version of the EMA \cite{barrerasPbSe} and some variations as proposed in \cite{barrerasPbSe} and \cite{epsilonPbSe}. Under our approximation, and using the effective masses and potentials barriers described before, the obtained  $E_{gap}$ for PbSe Qds are presented in Fig.~\ref{EgPbSe(R)}.
\\
As can be viewed from Fig.~\ref{EgPbSe(R)}, the proposed model (finite EMA approach) is close to the results obtained experimentally and using other approaches, reflecting that the parameters used to describe the PbSe Qd (effective masses and barriers) are correct. As we have done before, an estimation of the  energy shift due to the electron-hole Coulomb interaction $E_{e-h}$ can be calculated, using the first allowed transition for a PbSe Qd of $R=2nm$. We obtain $E_{e-h}\approx -0.03  eV$ that agrees with results presented previously \cite{Cheng}.

%\bibliographystyle{unsrt}
%\bibliography{refs}{}

\end{document}